# Status of the physics validation studies using Geant4 in ATLAS

On behalf of the ATLAS Geant4 Validation Team

A. Dell'Acqua
*CERN EP/SFT, Geneva, CH*
dellacqu@mail.cern.ch

The new simulation for the ATLAS detector at LHC is performed using Geant4 in a complete OO/C++ environment. In this framework the simulation of the various test beams for the different ATLAS subdetectors offers an excellent opportunity to perform physics validation studies over a wide range of physics domains: the electromagnetic processes, the individual hadronic interactions, the electromagnetic and hadronic signals in calorimeters. The simulation is implemented by paying special attention to all details of the experimental layout and by testing all possible physics processes which may be of relevance to the specific detector under test: the resulting simulation programs are often more detailed than the corresponding Geant3-based simulation suites. In this paper we present relevant features of muon, electron and pion signals in various ATLAS detectors. All remaining discrepancies between Geant4 and test-beam data are currently being addressed and progress is continuous. This work shows that Geant4 is becoming a mature and useful product for simulating specific features of large-scale detector systems.

## 1. INTRODUCTION

This work is a brief review of the results obtained in more than two years of activity in the field of the simulation in Geant4 by a large team operating in all sectors of the ATLAS experiment. We will summarize the strategies adopted for the Geant4 physics validation in ATLAS. We will also report on the main results obtained in the studies of muon energy loss and secondary production in the ATLAS calorimeters and in muon detectors. We then review the studies about the electromagnetic processes in tracking detectors and shower simulations in calorimeters. Hadronic interactions in tracking devices and calorimeters are also presented.

## 2. STRATEGIES FOR PHYSICS VALIDATIONS IN ATLAS

### 2.1. Geant4 physics benchmarking and validation

The features of interaction models in Geant4 are compared with similar features in Geant3.21 (here considered as the baseline for detector simulation) including variables not accessible in the experiment.

The differences in applied models, like the effect of cuts on simulation parameters in the different variable space (e.g. range cut vs. energy threshold) are also presented.

We used available experimental references from test beams for various sub-detectors and particle types to determine the prediction power of models in Geant4 (and Geant3) to estimate the Geant4 performance. Different subdetectors are sensitive to different effects (energy loss, track multiplicity, shower shape) and can be used for a better insight into the simulation models.

We tuned Geant4 models ("physics lists") and parameters (e.g. range cuts) for an optimal representation of the experimental detector signal with all relevant respects.

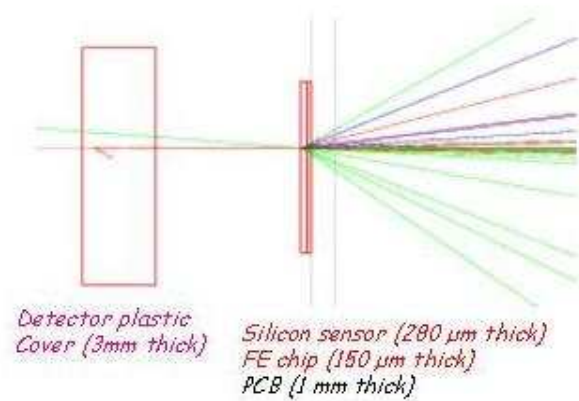

Fig. 1 – Hadronic interaction of a pion in an ATLAS silicon pixel module

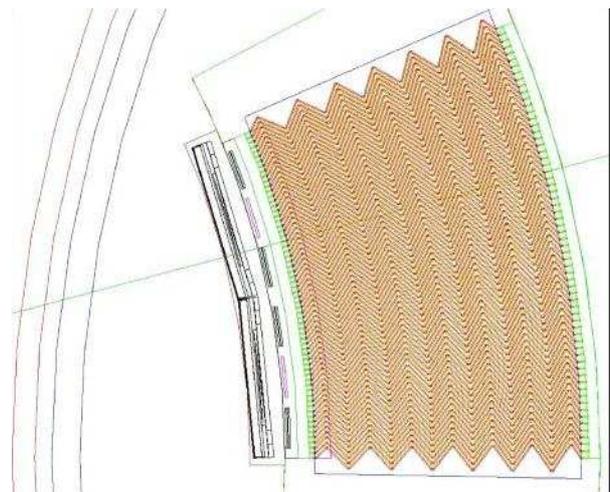

Fig. 2 – The electromagnetic barrel (Accordion) calorimeter

**MOMT003**

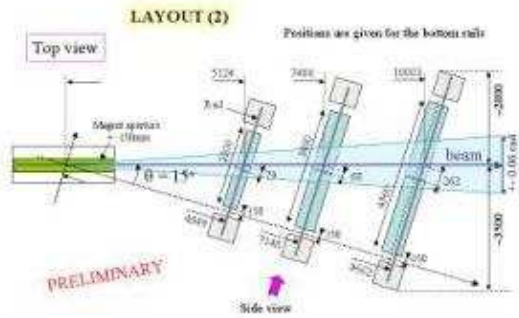

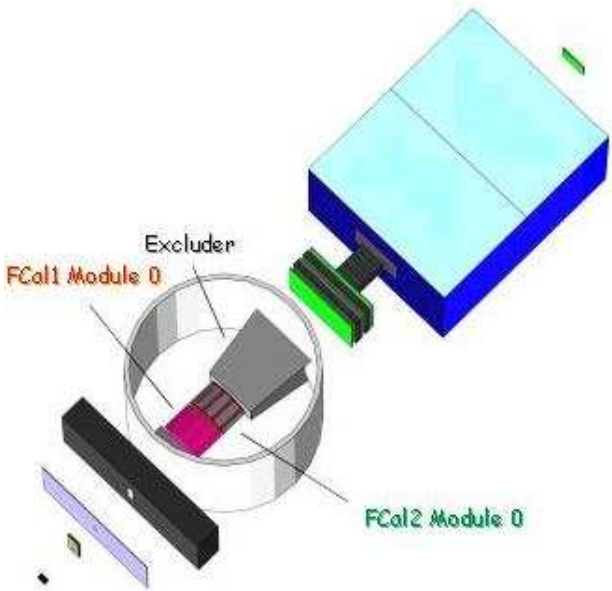

Fig.3 – The muon barrel test beam layout (CAD drawings and simulated setup)

Fig.4 – The Forward Calorimeter test beam setup (FCAL)

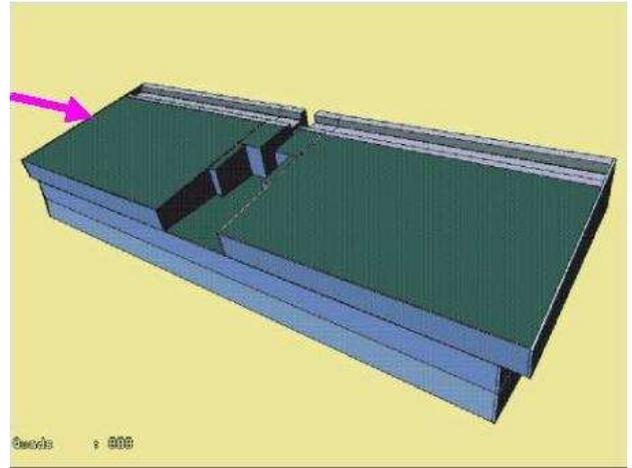

Fig. 5 – The tile calorimeter test beam layout (2000)

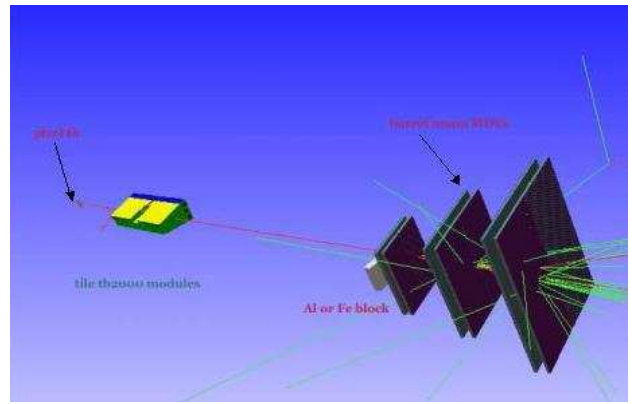

Fig.6 - The combined test beam setup (silicon pixel detector (left),tile calorimeter (center), muon barrel setup(right).

### 2.2. Geant4 validation strategies

The geometry description in the simulation should be as close as possible to the real test beam setup (active detectors and relevant elements of the experimental area, like magnets and inactive material in the beam). We also required the geometry descriptions in Geant3 and in Geant4 to be as close as possible when comparisons were to be made. To that purpose, we tried to use common databases or parameter books, as in the case of the muon detectors and calorimeters.

We generated particles in the simulation trying to reproduce as much as possible the real beam profile (e.g. in muon detectors and in calorimeters) and the momentum/energy distribution in the test beam: when needed, we tried to reproduce effects like beam contamination etc. The electronic readout features which can not be unfolded from the experimental signal were modeled in the simulation (coherent and incoherent electronic noise, digitization effect on the signal, to give some examples). Some of the ATLAS test beam setups which were simulated Geant4 are shown in Fig. 1,2,3,4,5 (the ATLAS pixel detectors, LAr electromagnetic calorimeter, muon system, forward calorimeter, tile calorimeter and the combined setup respectively.

## 3. MUON PHYSICS

### 3.1. Muon energy loss

The study of the energy lost by muons was performed in the hadronic endcap calorimeter (HEC) made of Liquid Argon/ Copper parallel plates. The experimental signal distribution is reproduced quite well by Geant4. Fig.7 shows the calorimeter signal distribution for 180 GeV muons for a .2mm range cut. The Geant4 simulation includes also a electronic noise simulation.

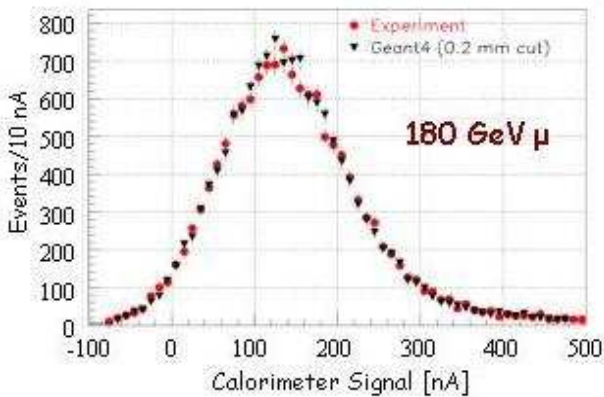

Fig. 7 – The hadronic endcap calorimeter signal.

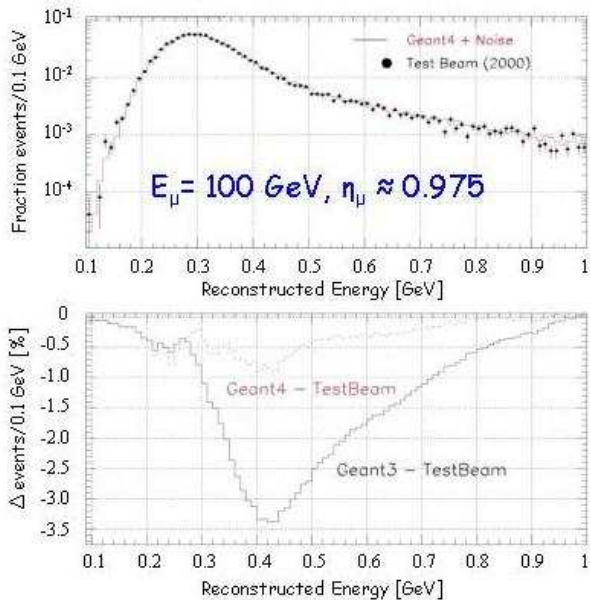

Fig.8 – Reconstructed energy (GeV) in the electromagnetic barrel calorimeter cells and comparison Geant3-data with Geant4-data for the same variable.

Some range cut dependence of Geant4 signal due to contribution from electromagnetic halo (δ-electrons) was observed, also in the case of tile calorimeter. The Geant4/test beam data comparison was performed also in the case of the electromagnetic barrel calorimeter (Liquid Argon/Lead in an accordion-like geometry) and a good agreement was observed in the distribution of reconstructed energy for incoming muons at 100 GeV incident energy. Comparisons between Geant3, Geant4 and the experimental data show that Geant4 is much better at reproducing the signal in the calorimeter than Geant3 and that the discrepancy between data and simulation is well within 1% for all energies (Fig. 8).

### 3.2. Secondary production by muons

In the case of the muon detectors, during the data taking period of summer 2002, the effect of dead material in front or between the muon chambers (production of extra-hits) was studied by positioning Aluminum or Iron targets (10,20 and 30 cm thick) about 37 cm from the first muon chamber or between the chambers. The probability of extra hits was measured at various muon energies (20 and 200 GeV). Results show that Geant4 can reproduce the distance of a extra hit to the muon track quite well. After a detailed simulation and reconstruction of simulated beams, a comparison between reconstructed track segments from simulation and reconstructed experimental data shows an agreement which is well within 1% in the case of Al or Fe in front of the muon system setup.

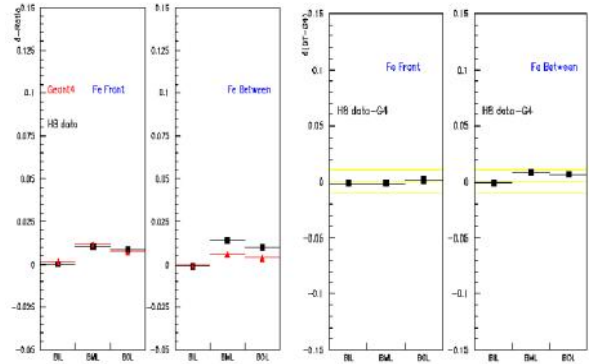

Fig.9 – Reconstructed track segments from simulation and experimental data comparisons for the muon test beam with dead material (Al or Fe) in the setup.

## 4. ELECTRON PHYSICS

### 4.1. Silicon detectors: ionization and PAI model

The standard ionisation model was compared to PAI model for 100 GeV pions crossing a Pixel detector module (280 mm thick silicon).

As shown in Fig. 10 the distributions around the peak are identical. The PAI model does not seem to link properly to δ-ray production but (more important in the case of ATLAS) the spatial distribution of the produced δ-rays is correct: to this effect, it has been found that the cut in range should match the desired detector resolution (10μm or less in the case of the ATLAS pixel detector).

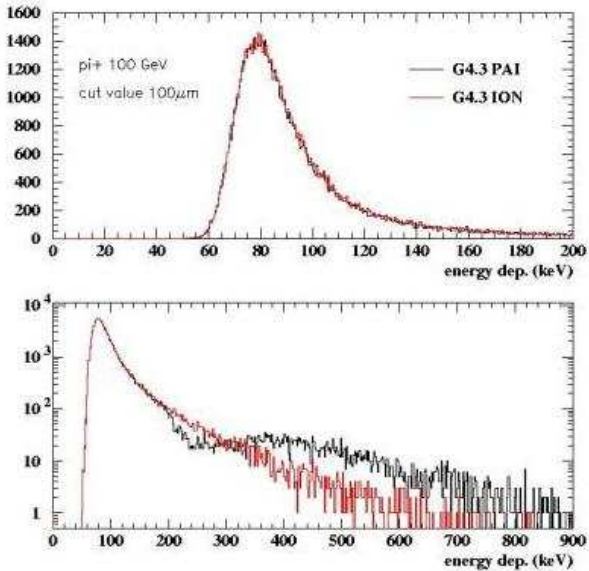

Fig.10 – Energy deposition in Silicon detectors (Pixel module) (keV) for a 100 GeV pion beam for the PAI model and the standard ionization model

### 4.2. Transition radiation detector

Very good agreement of Geant4 with data (and Geant3) for pions and muons beams was found. Several models were developed and tried for describing transition radiation, but none of them could accurately reproduce the additional energy deposited in the straws.

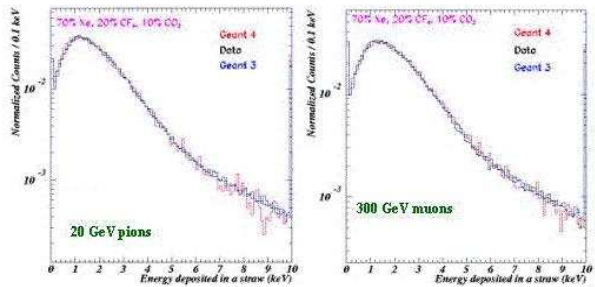

Fig. 11 – Energy distribution in the straws of the ATLAS TR tracker for pion and muon beams. Geant4 agrees very well with the experimental data (and Geant3)

Fig. 11 shows the distribution of the energy deposited in the straws by pions and muons (energy in keV): both Geant3 and Geant4 reproduce the experimental data accurately. Fig. 12 shows the energy deposited in the straws by electrons: when radiator foils are added in front of the straw planes, the contribution coming from transition radiation becomes obvious as a "shoulder" in the energy distribution, just above the 5keV threshold. Whilst the effect from transition radiation in the simulation is also noticeable, Geat4 does not reproduce the experimental data accurately. Moreover the use of transition radiation models in Geant4 has rather heavy repercussions on performance and turns out to be too demanding in terms of geometry and tracking. Hence it has been decided to suspend any further study on the subject and to convert an existing "home grown" TR model to be used from within Geant4.

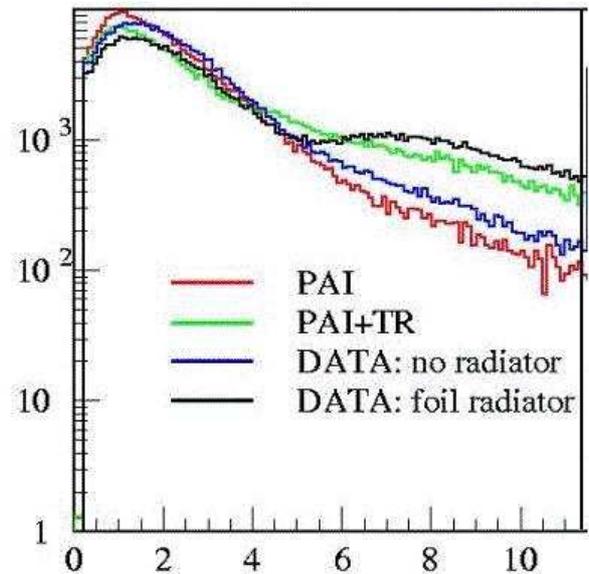

Fig. 12 – Energy distribution in the straws, in the case of an electron beam. The contribution from transition radiation is clearly visible as a "shoulder" above the 5keV threshold. The Geant4 simulation (in green) can not reproduce exactly the experimental data

### 4.3. Electron response in calorimeters

The characteristics of the electromagnetic showers in te ATLAS calorimeters have been studied quite accurately and compared with existing test beam data. In general terms, we find that Geant4 reproduces the average electron signal in all ATLAS calorimeters quite well. In the case of the electromagnetic barrel calorimeter ("Accordion") the agreement between simulation and experimental data is very good (Fig. 13). Energy fluctuations are reproduced quite well and the shower shape reproduces exactly what we find from the experimental data: a little discrepancy in the amount of energy deposited in the last longitudinal sampling is being addressed. The energy resolution for the ATLAS tile calorimeter is reproduced fairly well (Fig. 14), while by comparing the shower shape with what one gets from Geant3 we observe that electromagnetic showers tend to start earlier and to be more compact in Geant4 than in Geant3 (Fig.15).

Simulations of the electromagnetic compartment of the Forward calorimeter provide a constant term which, in the case of Geant4 is ~5%, to be compared with ~4% that one gets from the experimental data (and Geant3): this discrepancy is also being addressed.

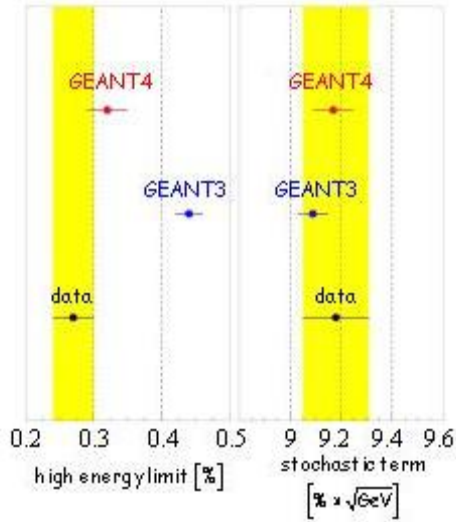

Fig. 13 – Energy resolution figures for the ATLAS Accordion calorimeter. Geant4 reproduces both the stochastic and the constant term quite accurately.

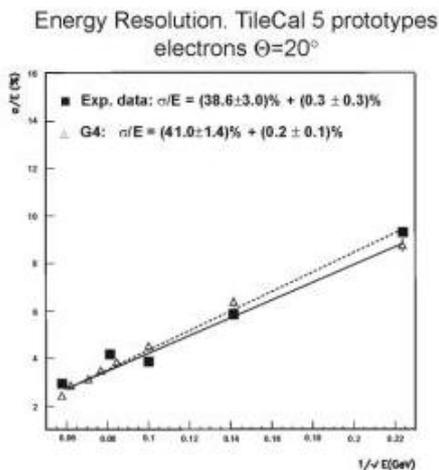

Fig. 14 – Energy resolution for electrons in the ATLAS tile calorimeter. The constant term is reproduced quite well, while the stochastic term from the simulation matches the experimental one within one standard deviation

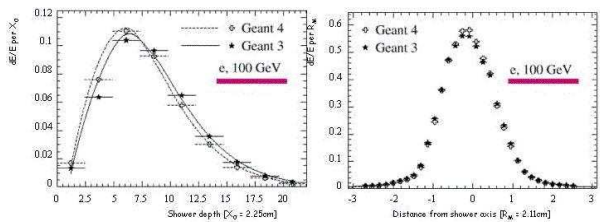

Fig. 15 – Shower shape in the ATLAS tile calorimeter. The shower starts earlier in Geant4 than in Geant3 and it is more compact

## 5. HADRONIC PHYSICS

### 5.1. Inelastic interactions in the pixel detectors

The energy from nuclear break-up released in hadronic inelastic interactions can cause large signals if a pixel (40μm X 500μm) is directly hits: this effect gives direct access to tests of single hadronic interactions, especially for what concerns the nuclear part. The parametric ("à la GHEISHA") model and the Quark Gluon String model in Geant4 have been compared to existing test beam data. The distributions of the energy released in the pixel detector show discrepancies between data and simulation, both in shape and average value (Fig.16 and 17). The parametric model fails to reproduce the fraction of energy deposited in one single pixel (the average value is 26% too small)

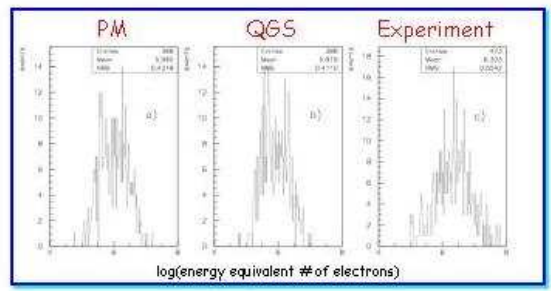

Fig. 16 – Energy deposited in the ATLAS pixel detectors by 180 GeV pions, from experimental data and from two different hadronic models in Geant4

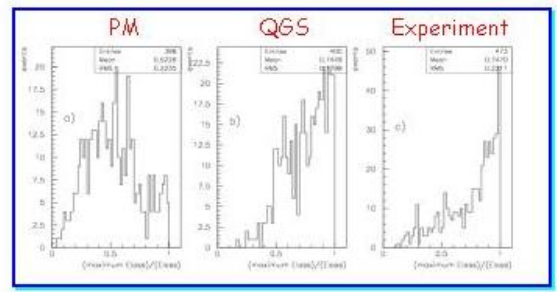

Fig. 17 – Fraction of energy deposited in one single pixel from experimental data and from two different hadronic models in Geant4

### 5.2. Hadronic physics with the ATLAS hadron calorimeters

Initial attempts of describing hadronic interactions in the ATLAS calorimeters provided rather poor results, due in part to the inadequacy of the hadronic models in Geant4 (parametric model, "à la GHEISHA") and in part to problems in matching low energy and high energy charged pion models, which produced "bumps" in the energy resolution distributions for all calorimeter types. The performance in simulating hadronic

6        UC*San Diego,CHEP03 , 23-27 March 2003*

interactions with Geant4 increased substantially with the introduction of new theoretical models (quark-gluon string models, QGS) and of new, better tuned parametric models (LHEP). The QGS models in particular seem to provide correct answers for all hadron calorimeters in ATLAS and show definite improvements over the initial parametric model and with respect to Geant3 (Fig. 18 and 19).

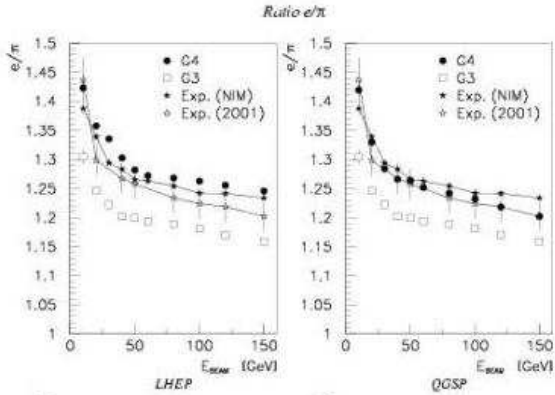

Fig. 18 – e/π ratio for the hadronic endcap calorimeter, for two Geant4 hadronic models. The QGS models seems to reproduce quite accurately the experimental data and performs much better than Geant3

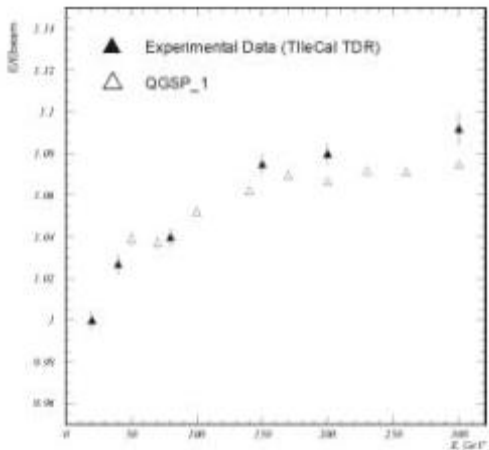

Fig. 19 – Pion non-linearity in the ATLAS tile calorimeter. The QGS models reproduce the experimental data quite accurately

The pion energy resolution is quite well reproduced by the QGS models in the case of the hadronic endcap calorimeter (Fig. 20): good results are also obtained for the tile calorimeter (Fig. 21).

A problem which is still under investigation is shown in Fig. 22, where the fraction of energy deposited in the four longitudinal samplings of the hadronic endcap calorimeter is plotted, for Geant3, Geant4 and the experimental data. It is apparent that in the current version of the QSG model, hadronic showers start too early , so that too much energy is deposited in the first two compartments and, correspondingly, too little is found in the following two: in this case Geant3 does a better job at reproducing the experimental data. The problem is under investigation and a fix is foreseen before too long.

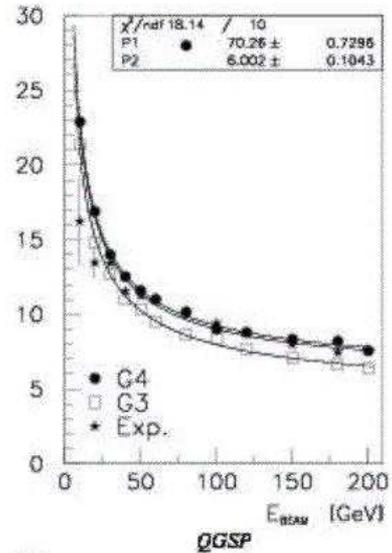

Fig 20 – Resolution curve for the ATLAS hadronic endcap calorimeter. Geant4 reproduces almost exactly the experimental data

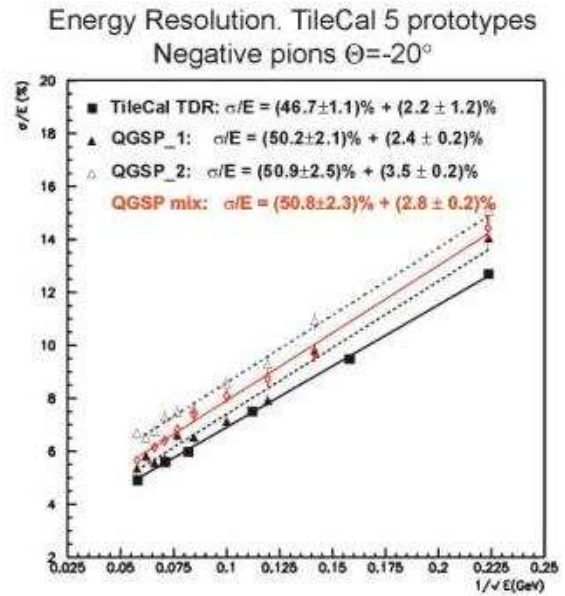

Fig 21 – Resolution curve for the ATLAS tile calorimeter. Geant4 (red line) reproduces almost exactly the constant term while the stochastic terms agree within the experimental errors

**MOMT003**

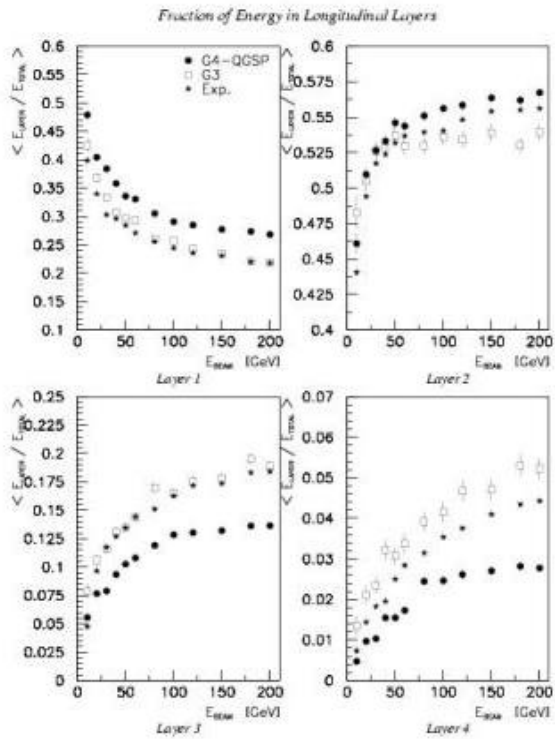

Fig. 22 – Fraction of energy deposited in the four longitudinal compartments of the ATLAS hadronic endcap calorimeter. Geant4 showers start too early and deposit too little energy in the last two compartments

## 6. CONCLUSIONS

Geant4 has been thoroughly tested in ATLAS in the last two years in order to evaluate its effective capability of reproducing and predicting the detector behaviour in the complex LHC environment.

All ATLAS subdetectors (and test beams) have been simulated and the detector's characteristics and sensitivities have been exploited. After an initial learning phase Geant4 was routinely used to detector and test beam simulations and has proven to be a rather rugged and dependable tool, which performs quite well and which provides in almost all cases a better detector simulation than Geant3.

It must be stressed that a good and solid collaboration was established since the beginning with the Geant4 team who demonstrated to be quite helpful in fixing and overcoming problems.

Except for a few open questions which are being addressed, there is general consensus within ATLAS that Geant4 is becoming a mature detector simulation tool that can be used for the challenging task of simulating a complex LHC detector.